\documentclass[12pt]{iopart}

\usepackage{amsfonts,mathrsfs,amsthm,amssymb}
\usepackage{graphicx,epsfig}
\usepackage[colorlinks=true,breaklinks=true,linkcolor=blue,citecolor=blue]{hyperref}

\def \ket {\rangle}
\def \bra {\langle}

\makeatletter
    
    \newcommand{\Rmnum}[1]{\expandafter\@slowromancap\romannumeral #1@}
\makeatother

\begin{document}
\title{Influence of detector motion on discrimination between photon polarizations}
\author{Tao Zhou,$^{1,2}$\footnote{zhoutao08@mails.tsinghua.edu.cn} Jingxin Cui,$^1$ and Ye Cao$^1$}
\address{$^1$State Key Laboratory of Low-dimensional Quantum Physics and Department of Physics, Tsinghua University, 100084 Beijing, China}
\address{$^2$Department of Physics, Sichuan University, 610064 Chengdu, China}

\date{\today}

\begin{abstract}
We investigate the discrimination between photon polarizations when measured by moving detectors. Both unambiguous and minimum-error discriminations are considered, and we analyze the the optimal successful (correct) probability as a function of the apparatus' velocity. The Holevo bound for polarization discrimination is also discussed and explicit calculation shows that the Holevo bound and the optimal successful (correct) probability for unambiguous (minimum-error) discrimination simultaneously increase or decrease.
\end{abstract}
\maketitle

\section{Introduction}
As a recent development, the possibility of discrimination between
quantum states can be potentially useful for many applications in
quantum computation and quantum communication.  In this problem, a
quantum state is chosen from a set of known states but we do not
know which and want to determine the actual states. If the states in
the set are not orthogonal, it cannot be successfully identified
with unit probability because of the non-cloning theorem. Two basic
strategies have been introduced to achieve the state discrimination,
one of which is the minimum-error discrimination \cite{Helstrom,Holevo,Yuen,Barnett,Andersson,Chou,Herzog} and
the other is the unambiguous discrimination
\cite{Dieks,Ivanovic,PeresPLA,Jaeger,Chefles,PeresJPA,Duan,Sun,CheflesPLA,PRA062107,Jafarizadeh,Wu}.
In the minimum-error discrimination, errors are permitted and the
optimum measurement is required such that the probability of error
is minimum. In the unambiguous discrimination not errors but
inconclusive results are permitted, and  in the optimum strategy the
probability of failure is a minimum. 

Recently, testing quantum mechanics for large space distances and eventually in implementing quantum information protocols in global scales attract a lot of interest~\cite{Aspelmeyer,Resch,Ursin,FedrizziNature}. Photon is an ideal physical object in quantum communications. Because of the present limits on the use of fiber optics in long distance communications, the most feasible alternative may be free-space transmission using satellites and ground stations. And then theoretical studies on the influence of the detector velocity are demanded by using satellites in quantum information experiments. In this paper, we address this issue by considering the discrimination between two photon polarizations when the measurements performed in different inertial frames are allowed. Pure polarization states for two monochromatic photons with different momenta can unambiguously distinguished in moving frames, while the polarizations of two non-monochromatic
photons cannot. Following the proposals in Ref.
\cite{PeresJMO}, the effective reduced density matrix for the
polarizations can be defined and calculated in moving frames, and
the polarization states can be distinguished with minimum error.

The organization of the paper is as follows. In Sec. \ref{sec2}, we
give a brief description of basis transformation under the Lorentz
boost. In Sec. \ref{sec3}, we discuss the discrimination between two pure polarizations of  two monochromatic photons in moving frames. How to calculate the effect reduced density matrix for photon polarizations is discussed in Sec. \ref{sec4}, and numeric results are shown for the minimum-error discrimination between polarizations of two non-monochromatic photons. In Sec. \ref{sec5}, we compare the Holevo bound and polarization discrimination. Finally, the paper is ended with a short discussion in Sec. \ref{sec6}.

\section{Relativistic state transformations for photons}\label{sec2}
To give the state transformation in different frames, we should
first discuss the photon basis states. We define the standard vector
$|\tilde{k},\sigma\ket$, where $\tilde{k}=(1,0,0,1)$, as follows
\begin{eqnarray}
P^\mu|\tilde{k},\sigma\ket&=&k^\mu|\tilde{k},\sigma\ket,\nonumber\\
J_z|\tilde{k},\sigma\ket&=&\sigma|\tilde{k},\sigma\ket,
\end{eqnarray}
and for photons $\sigma=\pm1$. The momentum-helicity eigenstates can
be generated from the standard vector $|\tilde{k},\sigma\ket$,
\begin{eqnarray}
|k,\sigma\ket=U(L_k)|\tilde{k},\sigma\ket,
\end{eqnarray}
where $k=L_k\tilde{k}$ is a four-component null vector, $k^2=0$ and
the helicity is denoted by $\sigma$. The choice of Lorentz
transformation $L_k$ is not unique \cite{GingrichPRA} and in the
present paper we set
\begin{eqnarray}
L_k=R(\hat{\mathbf{k}})L_z(k_0),
\end{eqnarray}
where $L_z(k_0)$ is a pure Lorentz boost along $z$ axis taking
$\tilde{k}$ to $k_0\tilde{k}$ and $R(\hat{\mathbf{k}})$ denotes a
rotation taking the vector $(1,0,0,1)$ to the vector
$(1,\hat{\mathbf{k}})$. In polar coordinate,
$\hat{\mathbf{k}}=(\sin\theta\cos\phi,\sin\theta\sin\phi,\cos\theta)$,
and $R(\hat{\mathbf{k}})$ can be chosen as
\begin{eqnarray}\label{eq5}
R(\hat{\mathbf{k}})=R_z(\phi)R_y(\theta).
\end{eqnarray}
The carrier space $\mathcal{H}$ of the irreducible representation of
the Poincar\'e group for photons is spanned by the momentum-helicity
eigenstates $|k,\sigma\ket$ and the basis are normalized by
\begin{eqnarray}
\bra
k,\sigma|k',\sigma'\ket=(2\pi)^3(2k^0)\delta_{\sigma\sigma'}\delta^{(3)}(\mathbf{k}-\mathbf{k}').
\end{eqnarray}

A Lorentz boost $\Lambda$ will induce a unitary operator
$U(\Lambda)$ on the Hilbert space $\mathcal{H}$
\cite{Wigner,Weinberg},
\begin{eqnarray}\label{WR}
U(\Lambda)|k,\sigma\ket=U(L_{\Lambda
k})U\big(W(\Lambda,k)\big)|\tilde{k},\sigma\ket,
\end{eqnarray}
where the Wigner rotation $W(\Lambda,k)=L^{-1}_{\Lambda k}\Lambda
L_k$ is an element in the little group which leaves $\tilde{k}$
invariant. For massless particles, the little group is the $E(2)$
group and in Eq. (\ref{WR}) $W(\Lambda,k)$ is just a rotation or
translation in the $x$-$y$ plane. Since the helicity is not affected
by translations, only a rotation by an angle $\Theta(\Lambda,k)$ is
left, and then
\begin{eqnarray}
U(\Lambda)|k,\sigma\ket=e^{-i\sigma \Theta(\Lambda,k)}|\Lambda
k,\sigma\ket.
\end{eqnarray}
The angle $\Theta(\Lambda,k)$ is explicitly given in Ref.~\cite{GingrichPRA},
\begin{eqnarray}\label{eq1}
\Theta(\Lambda,k)=\left\{ \begin{array}{ll}
0 & :\Lambda=L_z(k_0)\\
0 & :\Lambda=R_z(\gamma),\hat{\mathbf{k}}\neq \hat{\mathbf{z}}\\
\gamma & :\Lambda=R_z(\gamma),\hat{\mathbf{k}}=\hat{\mathbf{z}}\\
\arg(B+iA) & :\Lambda=R_y(\gamma)
\end{array}
\right.
\end{eqnarray}
for different Lorentz transforms and momenta, where
\begin{eqnarray}
A=\sin\gamma\sin\phi,\ \ \ 
B=\sin\gamma\cos\theta+\cos\gamma\sin\theta\nonumber.
\end{eqnarray}
Since all the Lorentz boosts  can be constructed by  $L_z$, $R_y$
and $R_x$, and
\begin{eqnarray}
W(\Lambda^\prime\Lambda,k)=W(\Lambda^\prime,\Lambda k)W(\Lambda,k),
\end{eqnarray}
$\Theta(\Lambda,p)$ for all $\Lambda$ any momentum $k$ can be
calculated from Eq.~(\ref{eq1}).

Now, we can use the two helicity states as a basis for the polarization states. The four-vectors of helicity states corresponding to momentum $\mathbf{k}$ are given by
\begin{eqnarray}
\epsilon^\pm_{\mathbf{k}}=R(\hat{\mathbf{k}})\epsilon^\pm,
\end{eqnarray}
where $\epsilon^\pm$ is the helicity vectors corresponding to the standard basis states $|\tilde{k},\sigma\ket$ and $R(\hat{\mathbf{k}})$ is the rotation taking the standard space direction $(0,0,1)$ to $\hat{\mathbf{k}}$, given in Eq.~(\ref{eq5}).
The polarization state $|\alpha(\mathbf{k})\ket$ for a photon with
momentum $\mathbf{k}$ can be expressed as
\begin{eqnarray}
|\alpha(\textbf{k})\ket=\alpha_+(\mathbf{k})|\epsilon_\mathbf{k}^+\ket+\alpha_-(\mathbf{k})|\epsilon_\mathbf{k}^-\ket,
\end{eqnarray}
with $|\alpha_+(\mathbf{k})|^2+|\alpha_-(\mathbf{k})|^2=1$. 
And a generic one-photon state is given by a wave-package~\cite{Mandel}
\begin{eqnarray}
|\Psi\ket=\int
d\mu(\textbf{k})f(\textbf{k})|k,\alpha(\textbf{k})\ket,
\end{eqnarray}
normalized by $\int d\mu(\mathbf{k})|f(\mathbf{k})|^2=1$ with the
Lorentz-invariant measure
\begin{eqnarray}
d\mu(\mathbf{k})=\frac{d^3\mathbf{k}}{(2\pi)^32k^0}.
\end{eqnarray}
According to Eq. (\ref{eq1}), the transformation for the
polarization under Lorentz boost $\Lambda$ is
\begin{eqnarray}\label{trans}
D(\Lambda)|\alpha(\mathbf{k})\ket=R(\Lambda
\hat{\mathbf{k}})R_z(\Theta(\Lambda,\mathbf{k}))
R(\hat{\mathbf{k}})^{-1}|\alpha(\mathbf{k})\ket.
\end{eqnarray}
When the Lorentz boost is along the $z$ axis, it can be simplified
as
\begin{eqnarray}\label{eq2}
D(\Lambda)|\alpha(\mathbf{k})\ket=R(\Lambda
\hat{\mathbf{k}})R(\hat{\mathbf{k}})^{-1}|\alpha(\mathbf{k})\ket.
\end{eqnarray}

\section{Unambiguous discrimination in moving frames}\label{sec3}

The polarized photon is an essential tool for both quantum communication and quantum computation. In quantum communication, optical fibers are usually used, and the photons may be absorbed or depolarized owing to the fiber's imperfections. In some cases, such as communication with space stations, the photons must propagate and the beam then has a diffraction angle. These mean that the photons are usually not monochromatic and have a momentum distribution in quantum communication. In this section, for simplicity, we discuss the idea case, where the photons are monochromatic. It will be shown that  the photon polarizations can be unambiguous distinguished in this case. In the next section, we will discuss a more realistic case, where the photons are non-monochromatic.

Let us assume that Alice prepares a single photon in one of the two polarization states $|\alpha_1\ket$ and $|\alpha_2\ket$ with equal probabilities. Besides the polarization freedom, the photon also has momentum freedom, and we assume that the two polarizations have two momenta $|\mathbf{k}_1\ket$ and $|\mathbf{k_2}\ket$, respectively. In this case, the photon corresponds to plane wave pulse. The receiver Bob tries
to unambiguously distinguish the two polarization states. We
consider the effect of Bob's motion relative to Alice, with a
constant velocity $\textbf{v}$.
 For convenience, we restrict that $\textbf{v}$ is along $\mathbf{k}_1$. A coordinate system in Alice's rest
 frame can be selected such that $k_1=(1,0,0,1)$, $k_2=(1,\sin\vartheta,0,\cos\vartheta)$ and $\textbf{v}=(0,0,v)$.
 And we suppose the receiver Bob has an infinite flat detetcor parallel to $x$-$y$ plane. In this paper, $c=1$ and
 we choose the zero-components of the momenta to be unit, because according to Eq. (\ref{eq1}) and (\ref{eq2}),
  the transformations for polarizations are independent on the magnitude of the momentum $\mathbf{k}$.
   The Lorentz transformation yields momenta's new components in Bob's rest frame,
\begin{eqnarray}
k_1^\prime &=&\big(\gamma(1-v),0,0,\gamma(1-v)\big),\nonumber\\
k_2^\prime
&=&\big(\gamma(1-v\cos\vartheta),\sin\vartheta,0,\gamma(\cos\vartheta-v)\big),
\end{eqnarray}
with $\gamma=(1-v^2)^{-1/2}$. New unit vectors of momenta in Bob's
rest frame are $\hat{\mathbf{k}}_1^\prime=(0,0,1)$ and
$\hat{\mathbf{k}}_2^\prime=(\sin\vartheta',0,\cos\vartheta')$, where
\begin{eqnarray}
\sin\vartheta'=\frac{\sin\vartheta}{\gamma(1-v\cos\vartheta)}.
\end{eqnarray}
According to Eq. (\ref{eq2}), the new polarization states are
\begin{eqnarray}\label{eq4}
|\alpha_1^\prime\ket=|\alpha_1\ket,\ \ \
|\alpha_2^\prime\ket
=R(\hat{\textbf{k}}_2^\prime)R(\hat{\textbf{k}}_2)^{-1}|\alpha_2\ket\,
\end{eqnarray}

To unambiguously distinguish the two polarizations, the POVM
detection operators for the optimum discrimination should be given.
Let the elements of POVM be $\Pi_1$, corresponding to unambiguously
detecting $|\alpha'_1\ket$, $\Pi_2$, corresponding to unambiguously
detecting $|\alpha'_2\ket$ and $\Pi_0$, corresponding to
inconclusive result.  The condition of no errors requires that
\begin{eqnarray}\label{Pi1Pi2}
\Pi_1|\alpha_2'\ket=0,\ \Pi_2|\alpha'_1\ket=0,
\end{eqnarray}
and in addition, because the POVM exhausts all possibilities, it is
implied that
\begin{eqnarray}\label{Pi0}
\Pi_0=I-\Pi_1-\Pi_2.
\end{eqnarray}
The probabilities of successfully identifying the two polarization
states is
\begin{eqnarray}
P=\frac{1}{2}\bra
\alpha_1^\prime|\Pi_1|\alpha_1^\prime\ket+\frac{1}{2}\bra
\alpha_2^\prime|\Pi_2|\alpha_1^\prime\ket.
\end{eqnarray}
The optimal POVM detection operators satisfying Eqs. (\ref{Pi1Pi2})
and (\ref{Pi0})  are given in Ref. \cite{Wu}
\begin{eqnarray}
\Pi_1=\frac{2}{3}|\alpha_1^{\prime\bot}\ket\bra\alpha_1^{\prime\bot}|,\ \
\Pi_2=\frac{2}{3}|\alpha_2^{\prime\bot}\ket\bra\alpha_2^{\prime\bot}|,\ \ 
\Pi_0=I-\Pi_1-\Pi_2,
\end{eqnarray}
and the optimal successful probability is
\begin{eqnarray}\label{eq3}
P_\mathrm{opt}&=&1-|\bra\alpha'_1|\alpha'_2\ket
=1-|\bra
\alpha_1|R(\hat{\textbf{k}}_2^\prime)R(\hat{\textbf{k}}_2)^{-1}|\alpha_2\ket|.
\end{eqnarray}
$|\alpha_1^{\prime\bot}\ket$ and $|\alpha_2^{\prime\bot}\ket$ are
called reciprocal basis \cite{Wu} which lie in the space spanned by
$|\alpha'_1\ket$ and $|\alpha'_2\ket$, defined as
\begin{eqnarray}
\bra\alpha_i^{\prime\bot}|\alpha'_j\ket=t_i\delta_{ij}.
\end{eqnarray}

\begin{figure}
\begin{center}
\epsfig{figure=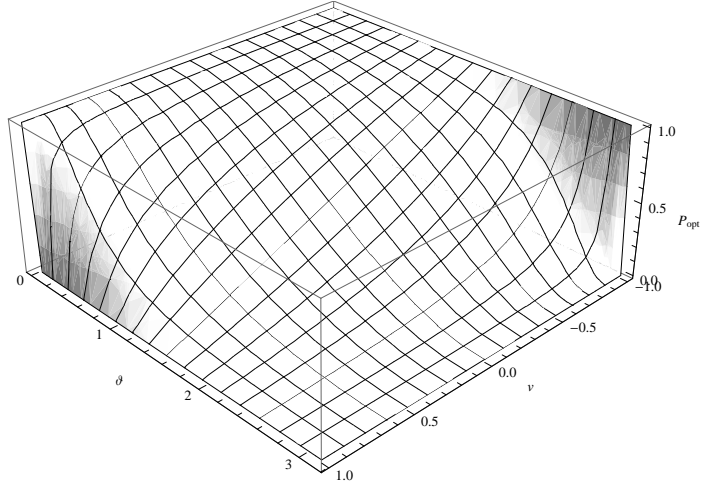,width=10cm}
\end{center}
\caption{The optimal successful possibility
$P_\mathrm{opt}$ in moving frames  as a function of $\vartheta$ and the
relative velocity $v$.}\label{f1}
\end{figure}

Eq. (\ref{eq3}) shows that the optimum successful probability
$P_\mathrm{opt}$ is dependent on the velocity $v$.  To make it more
obvious, we consider an example Alice prepares
$|\alpha_1\ket=|\epsilon_{\mathbf{k}_1}^+\ket$ and
$|\alpha_2\ket=|\epsilon_{\mathbf{k}_2}^-\ket$ with
$\mathbf{k}_1=(0,0,1)$ and
$\mathbf{k}_2=(\cos\vartheta,0,\sin\vartheta)$,
\begin{eqnarray}
|\epsilon_{\hat{\mathbf{k}}_1}^+\ket=\frac{1}{\sqrt{2}}\left(\begin{array}{c} 0\\
1 \\
i\\
0\end{array}\right),\ \ |\epsilon_{\mathbf{k}_2}^-\ket=\frac{1}{\sqrt{2}}\left(\begin{array}{c} 0\\
\cos\vartheta \\
-i\\
-\sin\vartheta\end{array}\right)
\end{eqnarray}
and then in Bob's rest frame
\begin{eqnarray}
|\alpha_1^\prime\ket=\frac{1}{\sqrt{2}}\left(\begin{array}{c} 0\\
1 \\
i\\
0\end{array}\right),\ \
|\alpha_2^\prime\ket=\frac{1}{\sqrt{2}}\left(\begin{array}{c} 0\\
\cos\vartheta' \\
-i\\
-\sin\vartheta'\end{array}\right)
\end{eqnarray}
and
\begin{eqnarray}\label{overlap}
\bra\alpha_1^\prime|\alpha_2^\prime\ket=\frac{1}{2}(\cos\vartheta'-1).
\end{eqnarray}
We see from Eq.~(\ref{overlap}) that the overlap between the two polarization states changes for different observers. This conclusion is reasonable since the transformation for photon polarization in different frames is dependent on the photon's momentum, which is shown in Eq.~(\ref{trans}).
Finally, the optimal successful possibilities are given as
\begin{eqnarray}
P_\mathrm{opt}(\vartheta,v)=\frac{(1+\cos\vartheta)(1-v)}{2(1-v\cos\vartheta)}.
\end{eqnarray}
The results are shown in Fig. \ref{f1}. When $\vartheta=0$ or
$\mathbf{k}_1=\mathbf{k}_2$, the optimum probabilities
$P_\mathrm{opt}$ are the same in different frames and there is no influence on the discrimination between the two polarizations, because the transformations for the two polarization states
 are the same. Howerver, if $\mathbf{k}_1\neq\mathbf{k}_2$, the optimal probability of unambiguous discrimination between the two polarization states is sensitive to Bob's relative velocity to Alice. It is obvious in Fig.~\ref{f1} that $P_{\mathrm{opt}}$ drops as $v$ increase. When Bob moves toward the opposite direction of z axis, with the magnitude of the velocity large enough ($v\rightarrow-1$), $P_{\mathrm{opt}}$ can become arbitrarily close to 1, which means the two polarization states can almost be perfectly distinguished even if Alice prepares two nonorthogonal polarization states. This shows how important the detector motion can be to polarization measurements when the velocity is high enough.

\section{Minimum-error discrimination in moving frames}\label{sec4}
As mentioned before, the photons in quantum communication are usually non-monochromatic, and in this section, we will discuss the more realistic case. Assume that Alice prepares  a single photon in one of the two helicity states. In the long-range propagation of the polarized photon, due to the imperfections of fiber or the diffraction in the free space, the photon may have a momentum distribution, and the corresponding one-photon states read
\begin{eqnarray}
|\Psi_\pm\ket=\int
d\mu(\textbf{k})f(\textbf{k})|\mathbf{k},\epsilon_\textbf{k}^\pm\ket,
\end{eqnarray}
where $f(\textbf{k})$ represents the momentum distribution. For simplify, we suppose that the momenta have a Gaussian distribution and the dispersion is restricted to $x-y$ plane,
\begin{eqnarray}
|f(\mathbf{k})|^2=\frac{1}{N}\exp(-\frac{k_r^2}{2\sigma^2})\delta(k_3-k_0),
\end{eqnarray}
where $k_r=\sqrt{k_1^2+k_2^2}$, and  $N$ is a normalization factor.
To discriminate between the two polarization states, we should first
calculate the reduced density matrix for the polarizations.
Following the proposals in Ref. \cite{PeresJMO}, an longitudinal
(unphysical) part of a polarization state $|\alpha_\mathbf{k}\ket$
can be defined as $\epsilon^l_\mathbf{k}=\hat{\mathbf{k}}$. A
polarization state along the $x$ axis is
\begin{eqnarray}
|\hat{\mathbf{x}}\ket=x_+(\mathbf{k})|\epsilon^+_{\mathbf{k}}\ket+x_-(\mathbf{k})|
\epsilon^-_{\mathbf{k}}\ket+x_l(\mathbf{k})|\epsilon^l_\mathbf{k}\ket.
\end{eqnarray}
Here,
$x_\pm(\mathbf{k})=\epsilon^\pm_\mathbf{k}\cdot\hat{\mathbf{x}}=(\cos\theta\cos\phi\pm
i\sin\phi)/\sqrt{2}$, and
$x_l(\mathbf{k})=\hat{\mathbf{x}}\cdot\hat{\mathbf{k}}=\sin\theta\cos\phi$.
The transverse part of $\hat{|\mathbf{x}\ket}$ is
\begin{eqnarray}
|b_x(\mathbf{k})\ket=x_+(\mathbf{k})|\epsilon^+_\mathbf{k}\ket+x_-(\mathbf{k})|\epsilon^-_\mathbf{k}\ket,
\end{eqnarray}
and similarly $|b_y(\mathbf{k})\ket$ and $|b_z(\mathbf{k})\ket$ can
be obtained. Then, we can define
\begin{eqnarray}
E_{mn}=\int
d\mu(\mathbf{k})|\mathbf{k},b_m(\mathbf{k})\ket\bra\mathbf{k},b_n(\mathbf{k})|,\
\ m,n=x,y,z.
\end{eqnarray}
Then, the effective reduced density matrix for polarization of a
one-photon state $|\Psi\ket$ can be expressed as
\begin{eqnarray}
\rho_{mn}&=&\bra\Psi|E_{mn}|\Psi\ket\nonumber=\int d\mu(\mathbf{k})|f(\mathbf{k})|^2\bra\alpha(\mathbf{k})|b_m(\mathbf{k})
\ket\bra b_n(\mathbf{k})|\alpha(\mathbf{k})\ket.
\end{eqnarray}
According to Eq.~(\ref{eq2}), in a moving frame with relative velocity $\mathbf{v}=(0,0,v)$, the reduced polarization density matrix can be obtained
\begin{eqnarray}\label{reduced}
\rho'_{mn}&=&\int d\mu(\mathbf{k})|f(\mathbf{k})|^2\bra R(\Lambda
\hat{\mathbf{k}})R(\hat{\mathbf{k}})^{-1}\alpha(\mathbf{k})|b_m(\mathbf{k})\ket\nonumber\\
&&\times\bra b_n(\mathbf{k})|R(\Lambda
\hat{\mathbf{k}})R(\hat{\mathbf{k}})^{-1}\alpha(\mathbf{k})\ket
\end{eqnarray}

Using Eq.~(\ref{reduced}), we can calculate the reduced density matrix $\rho_\pm$ for $|\Psi_\pm\ket$ in Bob's frame. And at this point, the above unambiguous state discrimination is not appropriate for this case because the two spaces supported by $\rho_+$ and $\rho_-$
are the same. But we can still distinguish them with the
minimum-error strategy. For minimum-error discrimination
inconclusive results do not occur, so that $\Pi_0=0$ and we require
that the probability of errors in the discrimination procedure is a
minimum.

The error probability can be expressed as
\begin{eqnarray}
P_E=\frac{1}{2}\mathrm{Tr}(\rho_+\Pi_-)+\frac{1}{2}\mathrm{Tr}(\rho_-\Pi_+)=\frac{1}{2}+\frac{1}{2}\mathrm{Tr}[(\rho_--\rho_+)\Pi_+].
\end{eqnarray}
Introducing the operator
$\Omega=\rho_--\rho_+=\sum_k\omega_k|\phi_k\ket\bra\phi_k|$, and it
is obvious that the minimum of the error probability is obtained
when    $\Pi_+$ is the projector onto those eigenstates
$|\phi_k\ket$ of $\Omega$ that belong to negative eigenvalues
$\omega_k$. The optimum detection operators therefore read
\begin{eqnarray}
\Pi_+^{\mathrm{opt}}=\sum_{k<k_0}|\phi_k\ket\bra\phi_k|,\ \
\Pi_-^{\mathrm{opt}}=\sum_{k\geqslant k_0}|\phi_k\ket\bra\phi_k|,
\end{eqnarray}
where $\omega_k<0$ for $1\leqslant k<k_0$ and $\omega_k\geqslant0$
for $k\geqslant k_0$. Clearly, the optimal minimum-error measurement
for discriminating between two quantum states is a von Neumann
measurement. The resulting minimum-error probability is
\cite{Helstrom}
\begin{eqnarray}\label{Pe}
P_E=\frac{1}{2}-\frac{1}{4}\mathrm{Tr}|\rho_+-\rho_-|.
\end{eqnarray}
For any operator $O$, the operator $|O|$ is defined as $(O^\dag
O)^{1/2}$.

\begin{figure}
\begin{center}
\epsfig{figure=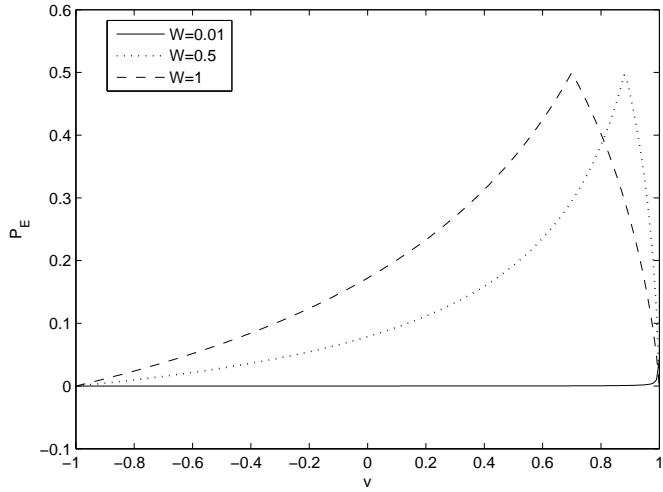,width=10cm}
\end{center}
\caption{The resulting minimum error probabilities
$P_E$ for distinguishing $\rho_+$ and $\rho_-$ in Bob's frame as a
function of the relative velocity $v$ between Alice and Bob.  Data
is shown for $W=0.01$, $W=0.5$ and $W=1$, where $W=\sigma/k_0$.}\label{f2}
\end{figure}

\begin{figure}
\begin{center}
\epsfig{figure=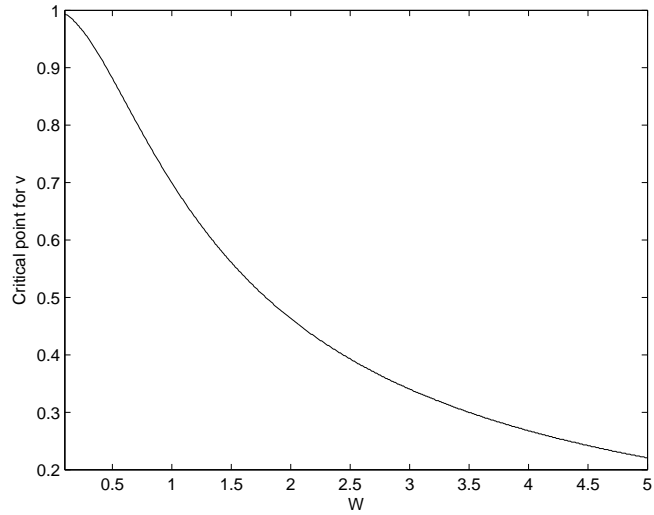,width=10cm}
\end{center}
\caption{The ``critical point'' for $v$ as a function
of wave packet width $W$.}\label{f3}
\end{figure}

Next, we perform a numerical investigation of Eq.~(\ref{Pe}).  Fig. \ref{f2} shows the minimum-error probability of discrimination
between $\rho_+$ and $\rho_-$ vs Bob's velocity $v$ relative to Alice, for $W=0.01$,  $W=0.5$, and  $W=1$, respectively. Here, $W$ is the wave packet width, and $W=\sigma/k_0$. We see that when Bob moves along the opposite direction of z axis, $P_E$ drops as $v$ decreases. However, there is a maximal value for $P_E$ when Bob moves along the same direction of z axis. This can be explained as follows. When $v=0$, the polar angles $\theta$ for different momenta satisfy $0\leqslant\theta\leqslant\pi/2$. $\theta$ drops as $v$ decreases and this helps to diminish $P_E$. And $\theta$ goes up as $v$ increases, and this will enlarge the minimum error probability $P_E$ before $\theta$ exceeds $\pi/2$. With $v$ large enough, $\theta$ will exceed $\pi/2$, this will lead to an opposite effect that helps to diminish $P_E$ again. Thus, we can define a ``critical point'' for the velocity
$v$, where the minimum error probability gets the maximal value. And
the ``critical point'' for $v$ vs $W$ is shown in Fig. \ref{f3}. The ``critical point'' decreases as $W$
goes up.

\section{Holevo bound and quantum state discrimination}\label{sec5}
Distinguishing quantum states is like gaining information. Alice has
a classic information source encoded in quantum states
$\rho_1,\rho_2,\ldots,\rho_n$, sent to Bob with probabilities
\{$p_1,p_2,\ldots,p_n$\}, and Bob tries to determinate the states to
obtain the information. The higher the successful probability is,
the more information Bob gets. A good measure of how much
information Bob can obtain is the accessible information. The upper
bound of accessible information called Holevo bound, is defined as
follows \cite{Nielson},
\begin{eqnarray}\label{HB}
\chi=S(\rho)-\sum_i p_iS(\rho_i),
\end{eqnarray}
where $\rho=\sum_ip_i\rho_i$ and $S(\rho)=-\mathrm{Tr}\rho\log\rho$ is
the von Neumann entropy for $\rho$. Since the von Neumann entropy
for photon polarization is not a relativistic scalar, which is
similar to that for massive particles \cite{PeresPRL,PeresRMP}, the
Holevo bound is not invariant in different frames. This is the
reason why the discrimination between photon polarization is influenced by detector motion. We compare the Holevo bound and photon polarization discrimination in the following.

\begin{figure}
\begin{center}
\epsfig{figure=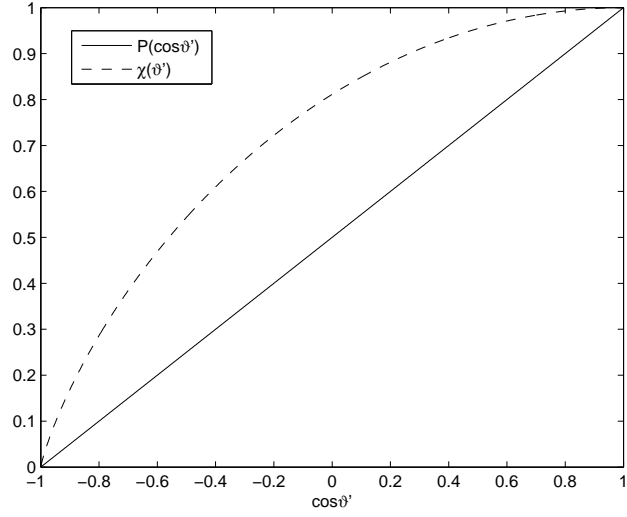,width=10cm}
\end{center}
\caption{The Holevo bound $\chi(\cos\vartheta')$ and
successful possibility $P(\cos\vartheta')$ are plotted as a function
of $\cos\vartheta'$. Both $\chi(\cos\vartheta')$ and
$P(\cos\vartheta')$ increase as $\cos\vartheta'$
increasing.}\label{f4}
\end{figure}
For the unambiguous discrimination between
$|\epsilon_{\textbf{k}_1}^+\ket$ and
$|\epsilon_{\textbf{k}_2}^-\ket$ above, it is easy to obtain the
Holevo bounds in moving frames. Simple calculation yields
\begin{eqnarray}
\chi(\cos\vartheta')&=&-\frac{1+\cos\vartheta'}{4}\log\frac{1+\cos\vartheta'}{4}-\frac{3-\cos\vartheta'}{4}\log\frac{3-\cos\vartheta'}{4},
\end{eqnarray}
where $\cos\vartheta'$ corresponds to the relative velocity $v$.
The Holevo bound $\chi(\cos\vartheta')$ and the successful probability
$P(\cos\vartheta')$ are shown in Fig. \ref{f4}. Both
$\chi(\cos\vartheta')$ and $P(\vartheta')$ increase as $\cos\vartheta'$ goes up.

\begin{figure}
\begin{center}
\epsfig{figure=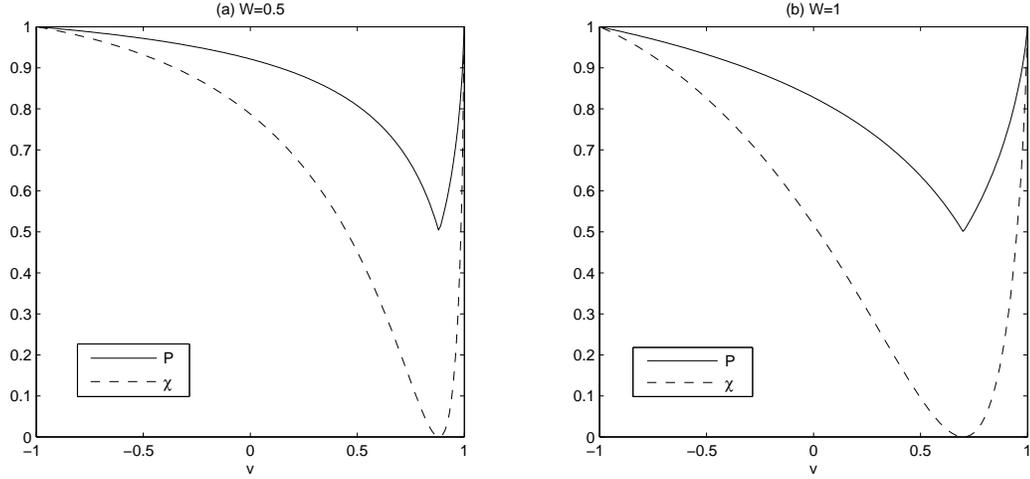,width=17cm}
\end{center}
\caption{The Holevo $\chi(\rho_+,\rho_-)$ and optimal
correct probability $P(\rho_+,\rho_-)$ are plotted as a function of velocity $v$, (a) for $W=0.5$ and (b) for $W=1$. $\chi(\rho_+,\rho_-)$ and $P(\rho_+,\rho_-)$
simultaneously decrease or increase as $v$ changes.}\label{f5}
\end{figure}

For the minimum-error discrimination between $\rho_+$ and $\rho_-$
that we mentioned earlier, the Holevo bound can be numerically given
according to Eqs. (\ref{reduced}) and (\ref{HB}). The Holevo bound
$\chi(\rho_+,\rho_-)$ and optimal correct possibility
$P(\rho_+,\rho_-)$ are shown in Fig. \ref{f5}, for $W=0.5$ and
$W=1$, respectively. When Bob moves along the opposite direction of z axis, $v<0$, both
$\chi(\rho_+,\rho_-)$ and $P(\rho_+,\rho_-)$ simultaneously increase
as $v$ decreases. When Bob move  along the same direction of z axis,
$\chi(\rho_+,\rho_-)$ and $P(\rho_+,\rho_-)$ decrease as the
magnitude of $v$ increases, and both reach the minimum values at the ``critical point'' of $v$. After the velocity $v$ exceeds the
``critical point'', $\chi(\rho_+,\rho_-)$ and $P(\rho_+,\rho_-)$
increase again with $v$ rising.

The two examples above reveal that, both the Holevo bound and the
optimal successful (or correct) probability simultaneously decrease
or increase as $v$ changes. The discrimination between photon polarizations is influenced by the measurement apparatus velocity.

\section{Conclusions and discussions}\label{sec6}
In summary, we investigate the influence of detector velocity on discrimination between photon polarizations. The successful (correct) probability for unambiguous (minimum-error discrimination)
is dependent on the apparatus velocity $v$ relative to the emitter.
For some cases, there are  ``critical points''  for the apparatus
velocity at which the correct probabilities to distinguish the polarozations reach the maximal
values. The Holevo bound and polarization discrimination are
also compared in the present work, and we discover that they
simultaneously decrease or increase in different frames.

\section*{Acknowledgments}
This work was supported by the National Fundamental Research Program
Grant Nos. 2009CB929402, 2011CB921602, and
 China National Natural Science Foundation
Grant No. 10874098.

\end{document}